\documentstyle[aps,epsf,preprint]{revtex}

\def\[{\left[}
\def\]{\right]}
\def\nn{\nonumber}
\def\({\left(}
\def\){\right)}
\def\labels#1{\label{#1}}
\def\eq#1{(\ref{#1})}
\def\d{\delta}

\def\l{\lambda}

\def\.{\cdot}

\def\.{\!\cdot\!}
\def\bi{\begin{itemize}}

\def\ei{\end{itemize}}
\def\be{\begin{eqnarray}}
\def\ee{\end{eqnarray}}
\def\bn{\begin{enumerate}}
\def\en{\end{enumerate}}
\def\h{{1\over 2}}

\def\nn{\nonumber}
\def\l{\lambda}

\def\r2{\sqrt{2}}

\def\x{\times}

\def\eq#1{(\ref{#1})}
\def\A{{\cal A}}

\def\t{\tau}

\def\d{\ \delta}

\def\rr2{{1\over\sqrt{2}}}

\begin{document}

\title{The Absorptive Extra Dimensions}
\author{D. Charuchittipan and C.S. Lam}
\address{Department of Physics, McGill University\\
3600 University St., Montreal, Q.C., Canada H3A 2T8\\
Email: Doojdao@physics.mcgill.ca, Lam@physics.mcgill.ca}
\maketitle

\begin{abstract}
It is well known that gravity and neutrino oscillation can be used to
probe large extra dimensions in a braneworld scenario.
We argue that neutrino oscillation remains
a useful probe even when the extra dimensions are small, because
the brane-bulk coupling is likely to be large. Neutrino oscillation 
in the presence of  a 
strong brane-bulk coupling is vastly different from the usual case of
a weak coupling. In particular, some active neutrinos could be absorbed
by the bulk when they oscillate from one kind to another, a signature which 
can be taken as the presence of an extra dimension.
In a very large
class of models which we shall discuss, the amount of 
absorption for all neutrino
oscillations is controlled by a single parameter, a property
which distinguishes extra dimensions 
from other mechanisms for losing neutrino fluxes.
\end{abstract}

\section{Introduction}
In the braneworld scenario \cite{BRANWORLD}, only gravity and sterile neutrinos
can move off our 3-brane into the bulk, so it is only these two kinds of
objects that can be used to probe the extra dimensions. If the size of the
extra dimensions is considerably smaller than 0.1 mm, or there is at
most one extra dimension with that size, gravity cannot be used 
to reveal the presence of an extra dimension,  so it is
up to the neutrinos. Assuming the coupling between the neutrinos in the brane and the bulk
to be large, we shall argue that
it may be possible to detect relatively small extra dimensions using precise neutrino 
oscillation data in the (distant) future.

The strength of this coupling is of course unknown, but in our minds it is
likely to be large. It is usually assumed that 
the coupling between the active brane neutrinos
and the Kaluza-Klein (KK) neutrinos of the bulk is of the Dirac-mass type. 
If this Dirac mass is comparable to the charged fermion masses, then the
brane-bulk coupling is indeed strong because the measure of its strength
is the ratio of the Dirac mass to the Majorana mass of
the neutrinos, a very large ratio.

This coupling is taken to be small in most of the
discussions in the 
literature \cite{LIT}. In that case the result follows from
perturbative calculations. 
Some, such as Barbieri et al, and Cosme et al \cite{LIT},
also considered the effect at large couplings via numerical simulations.
In  Ref.~\cite{LAM1,LAM2,EZAWA}, we studied the strong coupling limit
of a minimal model (MM) {\it analytically}. This is a five-dimensional model
with the smallest
number of parameters which can accommodate the present experimental
results on neutrino oscillations.
The details and the consequences, which will be explained more fully below and in the next section, 
turn out to be very 
different from what one can expect by extrapolation from the weak
coupling limit. In particular, to get three mass eigenstates on the brane
in the strong coupling limit, we have to start out with four brane neutrinos
when the coupling is turned off. The results are also distinct.
A small solar to atmospheric
mass-gap ratio naturally leads to a small reactor angle $\theta_{13}$, as observed. 
There is also an absorption of the active neutrino fluxes by the bulk. 
We shall argue below that this absorption, if experimentally observed
and if it satisfies the properties to be described later, is a good
indication of the presence of extra dimensions.

The purpose of this paper is to show that these distinct predictions
are generic, whose validity goes way beyond the MM. We shall produce a large class
of models, including those with a non-zero bulk masses, and those
with non-trivial bulk-bulk interactions, as well as some models with more
bulk neutrinos and/or in more
extra dimensions, which yields the same
essential properties as the MM, and gives rise to the same
predictions. This finding then allows us to accept the
consequences as a generic feature of the extra
dimensions, rather than some artifacts coming from the
specific assumptions of the MM.

We shall now define the MM and summarize its properties in the strong-coupling limit.
For more details please consult the next section.
The MM is a model in five spacetime dimensions, with a single massless
neutrino in the bulk. Its KK states have an integral mass spectrum, in an
energy unit which is inversely proportional to the size $R$
of the extra dimension. We shall adopt this unit throughout, so that all masses
are expressed as dimensionless parameters.
There are {\it four} left-handed neutrinos on the brane, 
the usual three, plus a sterile neutrino. 
They are assumed not to interact directly among themselves, but each
of the four will interact with the bulk neutrino with its own coupling
constant. In this model, the observed mixing between active neutrinos 
is induced from their individual couplings with the bulk.
The fourth neutrino
is forced on us by the strong coupling, because in that limit one 
brane neutrino always disappears into the KK towers. 
The details of how that comes about will be reviewed in the next section.
In order to have three 
mass eigenstates on the brane in the presence of a strong coupling, 
corresponding to the three active neutrinos with definite masses, we must start
from four brane neutrinos in the no-coupling limit. We shall refer to these four brane neutrinos {\it before
the coupling is turned on} as `flavor' neutrinos. Three of them are the usual active 
neutrinos, produced and absorbed by weak interactions in the usual way,
but of course not the fourth because it is sterile. In other words, the fourth flavor
is really flavorless. It is a {\it flavor}(-less) state only in the sense that a hadron is a {\it color}(-singlet) state.

A distinct feature of the strong coupling is the absorption of 
the active neutrino fluxes by the bulk. This absorption is the result of
a destructive interference of the infinite number of 
amplitudes oscillating through
the KK mass eigenstates. If the coupling were weak, only a few
nearby KK modes are reached. In that case  more complicated oscillatory
patterns (than those without the presence of any KK mode)
are setup, but the amplitude remains oscillatory in nature, in the sense
that it will return to its full flux at some appropriate time.
When the coupling gets stronger, more KK modes
are involved, the destructive interference between these modes
begins to take shape to dampen the oscillation amplitude over a finite
time period. This feature can be seen in the numerical simulations in
the paper of Barbieri et al and the paper of Cosme et al in Ref.~[2].
Finally, as the coupling strength becomes infinite, all KK modes
participate, destructive interference from these modes becomes complete, 
every hint of oscillation from these modes is wiped out, and the bulk becomes
purely absorptive.
Neutrino {\it oscillations} now
proceed completely through the three mass eigenstates on the brane, but their
amplitudes are now damped by the absorption from the bulk.  

There are nine possible oscillation channels, from any one of the three active
neutrinos to any other one. In the MM, the absorptions from these nine
channels are described by a {\it single} parameter $m_4$. In other words,
there are eight relations between the amount of absorptions in these nine
channels. This distinct feature of absorption can be used to separate this
mechanism of loss of amplitude from others, such as decay, as we shall
discuss later. For now, let us describe in more detail what this parameter
$m_4$ is, in what sense the MM is minimal, and what is meant by the
strong coupling limit.
This model 
starts with 8 real parameters, the four Majorana masses $m_a$ of the 
flavor brane neutrinos, and their four Dirac-mass couplings $d_a$
to the neutrino in the bulk. There is a single bulk neutrino without
a five-dimensional mass, whose KK energy spectrum consists of the
integers, in the unit of $1/R$.
The overall coupling strength is measured by $d^2=\sum_{a=1}^4d_a^2$. 
The strong coupling limit is taken to mean that this strength is much larger
than the other dimensionless parameters involved, namely, $d\gg m_a, e_a\equiv d_a/d$,
and 1. If the Dirac mass $d/R$ is similar to the electron mass, then since the
Majorana masses are small, we only require the Dirac mass to be much larger
than the KK mass gap. This condition
$d\gg 1$, with $d/R\simeq$ 0.5 MeV, is equivalent to
$R\gg 1/(0.5$ MeV$)\simeq$ 400 fm, so it remains to be in the strong coupling
limit even to fairly small
extra dimensions. In that limit, 
we are left with 7 parameters.
Six of them can be taken to be the active neutrino masses $M_1,M_2,M_3$,
and the mixing angles $\theta_{12},\theta_{23}$, and $\theta_{13}$.
The 7th parameter $m_4$, which is the Majorana mass of the sterile
neutrino before any coupling is turned on, measures the amount of absorption by
the bulk. Absorption is absent in the limit $m_4\to\infty$. 

We now argue that an experimental observation of these distinct absorption patterns 
is a good indication of the presence of extra dimensions. As discussed above,
absorption occurs only when many KK modes participate in a destructive 
interference. Since there is no simple reason to expect the presence of {\it many}
sterile neutrinos without an extra dimension, we think
that the presence of extra dimensions is strongly indicated if such an observation 
is made. However, partial loss of an oscillation amplitude may also come from decay \cite{DECAY},
oscillation into a very heavy neutrino \cite{HEAVY}, or other mechanisms. To distinguish
these other possibilities from the absorption by the bulk,   a quantitative analysis
is required. In the absorption mechanism, a single parameter $m_4$ governs the nine 
oscillation channels, so one can predict the amount of absorption of the other eight
channels when one is measured. If this is verified, then one can be reasonably sure
that the loss is due to the absorption in the bulk, and not something else. In other
words, the presence of an extra dimension is indicated.

In this paper we shall show that the results  of the MM
are rather generic. These results
include the control of all absorptions by a single parameter $m_4$,
which allows us to distinguish extra dimensional absorption from others.
They also include the natural connection between a small solar to atmospheric
mass-gap ratio and the smallness of the reactor angle $\theta_{13}$. 
We will show that these features
are preserved in a much larger class of models
in which the bulk spectrum is arbitrarily altered, so that the nature
and the details in the bulk are not important, only its presence is. 
This is so because the amount of absorption in the strong coupling
limit is controlled by the
same seven parameters in the MM, and not the details of the KK
spectrum.

In Sec.~2 we discuss the mass matrix of the MM, and its generalization.
The diagonalization of this infinite-dimensional matrix to find the mass
eigenvalues and the PMNS mixing matrix is worked out. These results are
used in Sec.~3 to compute the oscillation amplitudes and the absorption
by the bulk. A conclusion can be found in Sec.~4.
More complicated technical aspects are discussed in Appendix A.  Throughout
the rest of this article, we will not restrict the number of brane neutrinos
to $f=4$.

\section{The Strong-Coupling Models}
The mixing of four-dimensional neutrinos is given by a
(real symmetric) mass matrix ${\cal M}$. In the Minimal Model (MM), it is
\be
{\cal M}=\pmatrix{m&D\cr D^T&B\cr},\labels{mm}\ee
where $m={\rm diag}(m_1,m_2,\cdots,m_f)$ is the Majorana
mass matrix of the $f$ brane neutrinos.
$B={\rm diag}(0,+1,-1,+2,-2,\cdots)$ is the
mass matrix of the bulk neutrinos, which are the Kaluza-Klein
(KK) states of a single massless neutrino in a five-dimensional bulk. The
coupling between the brane and the bulk neutrinos
 is assumed to be of  the Dirac-mass type,
given by the  $f\x \infty$ 
matrix $D$, with $D_{an}=d_a$. The matrix elements of $D$ are independent
of $n\in{\bf Z}$ because every brane neutrino couples to the whole neutrino
in the bulk, so it couples equally to each of its KK states.

In this paper we generalize $B$ to an arbitrary real
symmetric matrix. In other words, we assume the spectrum of the bulk states
that couple to the brane neutrinos to be rather arbitrary. They no longer
need to come from a single massless neutrino in the bulk, nor do the
neutrinos in the bulk need to be restricted to a five-dimensional spacetime.
Some of the concrete models within this category will
be discussed in Sec.~3.

We shall use the discrete index $n$ to label the bulk states. In the MM,
$n\in{\bf Z}$, but in general, the set of $n$ can be quite complicated.
We shall denote it by ${\bf K}_0$.

The eigenvalue equation for the mass matrix is
\be
M\pmatrix{w\cr v\cr}=\l\pmatrix{w\cr v\cr},\labels{meigen}\ee
where $w$ is a $f$-dimensional column vector with components $w_a$, 
and $v$ is an $\infty$-dimensional column vector with components $v_n$. $\l$
is the eigenvalue. In component form, \eq{meigen} reads
\be
m_aw_a+d_aA&=&\l w_a,\labels{eiv1}\\
b+(Bv)_n&=&\l v_n,\labels{eiv2}\ee
where
\be
A&=&\sum_nv_n,\nn\\
b&=&\sum_{a=1}^fd_aw_a.\labels{b}\ee
We shall choose the normalization of the eigenvector so that $b=1$. 

Since $B$ is real symmetric, it can be diagonlized by a real orthogonal
matrix $O$, so that
\be
B=O^{-1}\.\mu\.O,\labels{diag}\ee
where $\mu$ is a diagonal matrix with real matrix elements $\mu_n$. 
In the MM, $O_{mn}=\d_{mn}$ and $\mu_n=n$.
Defining
$u=Ov$, and remembering the normalization $b=1$, \eq{eiv2} becomes
\be
\xi_n+\mu_nu_n=\l u_n,\labels{eiv3}\ee
where 
\be
\xi_m\equiv\sum_nO_{mn}.\labels{rho}\ee
The eigenvector components can be solved from \eq{eiv1} and \eq{eiv2} to be
\be
u_n&=&{\xi_n\over\l-\mu_n},\nn\\
w_a&=&A{d_a\over \l-m_a}=(Ad){e_a\over\l-m_a}\labels{eivec}\ee 
For later convenience we have expressed $d_a=de_a$, where
\be
d^2&=&\sum_{a=1}^fd_a^2,\nn\\
1&=&\sum_{a=1}^fe_a^2.\labels{e2}\ee
The constant $A$ may now be computed to be
\be
A=\sum_nv_n=\sum_{m,n}O_{mn}u_m=\sum_m{\xi_m^2\over\l-\mu_m},\labels{a}\ee
where orthogonality of the matrix $O$ has been used. 

The eigenvalue  
equation is obtained from \eq{b} and \eq{eivec}
and the normalization condition $b=1$ to be
\be
1=\sum_{a=1}^fd_aw_a=Ad^2\sum_{a=1}^f{e_a^2\over\l-m_a}\equiv Ad^2r.
\labels{eiv5}\ee
In the strong-coupling limit when $d\to\infty$, with 
the other parameters $e_a,m_a,\mu_k$ and $\xi_k$ kept fixed, 
eigenvalues either satisfy $A(\l)=0$, or $r(\l)=0$. We refer to the
former as bulk eigenvalues, and the latter as brane eigenvalues.

It is impossible to solve these eigenvalues analytically, especially
for large $f$, but we know enough about them to make a substantial progress.
Let us first consider
the brane eigenvalues determined by $r(\l)=0$. Since  $r(\l)$
goes to $+\infty$ at $\l=m_a+$ and  $-\infty$ at $m_a-$, 
it must cross zero somewhere between 
one $m_a$ and the next. Consequently
there are $f-1$ brane zeros all together, sandwiched
between consecutive pairs of $m_a$'s. We shall denote these
brane eigenvalues as $M_i$, with $i=1,2,\cdots,f-1$.
Note that these eigenvalues are completely independent of what $B$ is.
In particular, they are identical to those of the MM with the same
parameters $m_a$ and $d_a$.

The bulk eigenvalues behave similarly because $A(\l)$ has 
the same structure as
$r(\l)$. Hence there is one bulk eigenvalue 
between each consecutive pairs of $\mu_n$'s. The set of bulk eigenvalues
will be denoted by ${\bf K}$. In the MM, 
$\mu_n=n\in {\bf Z}={\bf K}_0$, and the bulk eigenvalues are ${\bf K}={\bf Z}+\h$.

We proceed to consider the mixing matrix $U$. Its columns are the 
normalized eigenvectors, with components
$U_{a\l}=w_a/N$ and $U_{n\l}=v_n/N$. The norm $N^2$ of the original eigenvector
$(w_a,v_n)$ is given by $N^2=(Ad)^2s+T$, where
\be
s(\l)&=&{1\over (Ad)^2}\sum_{a=1}^fw_a^2\nn\\
&=&\sum_{a=1}^f{e_a^2\over(\l-m_a)^2},\nn\\
T(\l)&=&\sum_nv_n^2=\sum_{n,k,\ell}O_{kn}u_kO_{\ell n}u_\ell\nn\\
&=&\sum_k{\xi_k^2\over
(\l-\mu_k)^2}.\labels{qt}\ee

If $d$ is large but not infinite, the eigenvalues will shift somewhat, but
they are still bounded between consecutive $m_a$'s or consecutive $\mu_n$'s.
The  
normalization factors in \eq{qt} have a $d$ dependence, to be
denoted by $s_d(\l)$ and $T_d(\l)$.
By definition, $s_\infty(\l)=s(\l)$ and $T_\infty(\l)=T(\l)$.

\section{Transition Amplitudes in the Absorptive Bulk}
Using the results obtained in the last section, 
we can calculate the transition amplitude $\A_{ab}$
from a brane neutrino of flavor $b$ and energy $E$,  to a brane neutrino of flavor $a$
after it has traversed a distance $L=2E\t/U^2$. 
The transition amplitude  is determined by the formula
\be
\A_{ab}(\t)&=&\sum_\l U^*_{a\l}U_{b\l}e^{-i\l^2\t}\equiv \A^S_{ab}(\t)
+\A^K_{ab}(\t),\labels{ta}\ee
where $\A^S$ is the contribution from the brane eigenvalues
$\l=M_i$, and $\A^K$ is the contribution from the bulk eigenvalues $\l\in {\bf K}$. We shall see that ${\cal A}^K_{ab}(\t)$ for $\t>0$ is zero
in the strong coupling limit, thus rendering the bulk an absorber of the brane neutrinos. This is the signature for the presence of an extra dimension.

When $d\to\infty$, the quantities 
$v_n, w_a/Ad, s$ and $T$ are all of order 1, so
the magnitude of $w_a$ is determined by $Ad$ and the magnitude of 
$N^2$ is determined by $(Ad)^2$. According to \eq{eiv5}, $Ad=1/(dr)$.
For bulk eigenvalues, $r=O(1)$, so $Ad=O(1/d)$. This implies $N^2\simeq T$
and $U_{a\l}=O(1/d)$. In that case the bulk components of an eigenvector
are much larger than the brane components. 
For brane eigenvalues, $A=O(1)$, hence $Ad=O(d)$ and $w_a=O(d)$. In that case
the brane components of an eigenvector dominate and $N^2\simeq (Ad)^2s$.

Let us denote the large-$d$ value of $U_{aM_i}$ by $V_{ai}$, and
$1/(\l_i-m_a)$ by $x_{ai}$. Then
\be
V_{ai}={e_ax_{ai}/\sqrt{s_i}}\quad(1\le a\le f,\ 1\le i\le f-1),\labels{vai}\ee
and
\be
\A^S_{ab}(\tau)=\sum_{i=1}^{f-1}V^*_{ai}V_{bi}e^{-iM_i^2\,\t}.\labels{asv}\ee
As it stands, $V$ is a $f\x(f-1)$ matrix, but we can make it into a
square $f\x f$ matrix by letting the last column to be $V_{af}=e_a$.
The meaning of this last column will be discussed later.
Note that we can write $V_{af}$ in the same form as the
other $V_{ai}$, namely, $V_{af}=e_ax_{af}/\sqrt{s_f}$, provided we
let $\l=\infty$.
The resulting square matrix
\be
V=\pmatrix{e_1x_{11}/\sqrt{s_1}&e_1x_{12}/\sqrt{s_2}&\cdots&
e_1x_{1,f-1}/\sqrt{s_{f-1}}&e_1\cr
e_2x_{21}/\sqrt{s_1}&e_2x_{22}/\sqrt{s_2}&\cdots&e_2x_{2,f-1}/\sqrt{s_{f-1}}
&e_2\cr
\cdots&\cdots&\cdots&\cdots&\cdots\cr
e_fx_{f1}/\sqrt{s_1}&e_fx_{f2}/\sqrt{s_2}&\cdots&e_fx_{f,f-1}/\sqrt{s_{f-1}}
&e_f\cr}\labels{v}\ee
is real orthogonal. This will be shown in Appendix A. 

Eq.~\eq{v} contains the parameters $e_a,m_a$, and $M_i$, but they are not 
all independent. The $f$ $e_a$'s are related by \eq{e2}, so only $f-1$
of them are independent. The brane eigenvalues $M_i$ are solutions of
$r(\l)=0$, so they are determined by the $2f-1$ parameters $e_a$ and $m_a$.
For large $f$, an analytic solution does not exist, so $M_i$ cannot
be expressed analytically in terms of $e_a$ and $m_a$, which makes 
any computation involving $M_i$, such as \eq{v} and \eq{asv}, difficult
to do in a closed form. To overcome this difficulty, it is crucial to make
the following observation.

Instead of the $f-1$ parameters $e_a$, we shall adopt the $f-1$ values
of $M_i$ as independent parameters. Then $e_a^2$ is a rational
 function of $M_i$ and $m_a$, as we shall show below. In this way
everything can be carried out analytically, and this fact is used heavily
in Appendix A to show the orthogonality of the matrix $V$.

The crucial point is the observation that $r(\l)$ is a meromorphic function
of $\l$, with $f$ simple poles at $\l=m_a$,  and $f-1$ zeros at $\l=M_i$.
Moreover, it follows from the definition of $r$ in \eq{eiv5} and the
constraint on $e_a$ in \eq{e2} that $r(\l)\to 1/\l$ for large $\l$. Hence
from complex variable theory we can conclude that
\be
r(\l)={\prod_{k=1}^{f-1}(\l-M_k)\over\prod_{c=1}^f(\l-m_c)}.\labels{rr}\ee
Since $e_a^2$ is the residue of $r(\l)$ at the simple pole $m_a$, it follows
that
\be
e_a^2=
{\prod_{k=1}^{f-1}(m_a-M_k)\over\prod_{c\not=a}^f(m_a-m_c)}\equiv
{\prod_{k=1}^{f-1}(-x_{ak})\over\prod_{c\not=a}^fm_{ac}},
\labels{e2res}\ee
where for later convenience we  introduce the abbreviations
\be
x_{ai}&=&M_i-m_a,\nn\\
m_{ab}&\equiv& m_a-m_b,\nn\\
M_{ij}&\equiv&M_i-M_j.\labels{mab}\ee
If there is one zero $M_i$ between each consecutive pair of poles, then
$e_a^2>0$. We shall refer to this as the {\it physical range}. Otherwise,
$e_a<0$ may occur.

Putting \eq{e2res} into \eq{qt}, we can compute $s_i=s(M_i)$. It is shown in
Appendix A that the sum over $a$ can be carried out to yield the simple result 
\be
s_i=-{\prod_{k\not=i}^{f-1}M_{ik}\over\prod_{c=1}^fx_{ci}}.\labels{qa}\ee

We consider now the contribution from the bulk eigenvalues. Since $U_{a\l}$
is unitary, it follows from \eq{ta} that $\A_{ab}(0)=\d_{ab}$, hence
\be
\A^K_{ab}(0)=\d_{ab}-\A^S_{ab}(0).\labels{ak0}\ee
Using \eq{asv} and the unitarity of the matrix $V$ in \eq{v}, we conclude
that
\be
\A^S_{ab}(0)=\d_{ab}-V_{af}^*V_{bf}=\d_{ab}-e_ae_f.\labels{as0}\ee
Therefore
\be
\A^K_{ab}(0)=e_ae_b.\labels{ak01}\ee

The contribution from the bulk eigenvalues can also be obtained directly
from \eq{ta} and the paragraph following that equation to be
\be
\A^K_{ab}(\t)&=&\sum_{\l\in {\bf K}}{1\over (dr)^2T_d(\l)} 
{e_ae_b\over(\l-m_a)(\l-m_b)}e^{-i\l^2\t}.\labels{aa}\ee
For a fixed $\l$, both $r$ and $T$ are of order 1 as 
$d\to\infty$, so the contribution
from each bulk eigenvalue to the sum is $O(1/d^2)$. Since there
is an infinite number of bulk eigenvalues, the total contribution to the
sum in \eq{aa} is not necessarily zero. In fact, we know 
from \eq{ak01} that $\A^K_{ab}(0)$ remains finite at infinite $d$. 

There is no harm in dropping a finite number of terms in the infinite sum since
each term contributes $O(1/d^2)$. For that reason we may assume $\l$
in the sum to be much larger than all the $m_a$'s, in which case
$r(\l)\simeq 1/\l^2$ because of \eq{e2}, so $(\l-m_a)(\l-m_b)r^2(\l)\simeq 1$.
The sum \eq{aa} can therefore be simplified to
\be
\A^K_{ab}(\t)&=&e_ae_b\sum_{\l\in {\bf K}}{1\over d^2T_d(\l)}\exp(-i\l^2\t)
\equiv e_ae_bg(\t),\labels{akg}\ee
where
\be
g(\t)&=&\sum_{\l\in {\bf K}}{1\over d^2T_d(\l)}\exp(-i\l^2\t).\labels{g}\ee
Since $g(0)=1$, we must have $\sum_{\l\in{\bf K}}T_d^{-1}(\l)\simeq d^2$
for large $d$. Coupling this with the earlier observation that $T_d(\l)=O(1)$
when $d\to\infty$, we conclude that $T^{-1}_d(\l)$, as a function of $\l$,
is of order 1 until
$|\l|=O(d^2)$, after which it decreases sufficiently fast in $\l$ for the
series to converge to $d^2$. Assuming $\mu_n$ to be smooth, and the
distances between one $\mu_n$ and another are bounded from above,
the additional oscillatory factor $\exp(-i\l^2\t)$ is going
to render $g(\t)\to 0$ for finite $\t$  when $d\to\infty$. 
This can be seen  by using the
natural variable $y=\l/d^2$. A very explicit calculation for the MM based on 
this reasoning can be found in Ref.~\cite{LAM1} and \cite{LAM2}.

In conclusion, the bulk contribution is 
\be
\A^K_{ab}(\t)=e_ae_bg(\t),\labels{akfinal}\ee
with $g(0)=1$ and $g(\t)\simeq 0$ for $\t\gg 1/d^4$.

\section{Conclusion}
In two previous publications \cite{LAM1,LAM2}, a five-dimensional minimal model (MM) of neutrino
oscillation was discussed. It is a model with a strong brane-bulk coupling. It contains seven parameters,
six of which can be taken to be the three masses of the active neutrinos and the three mixing angles.
The seventh, $m_4$, controls the absorption by the bulk of the oscillation amplitudes in the nine channels.
The MM model predicts the smallness of the solar to atmospheric mass-gap ratio and the smallness of the
reactor mixing angle are naturally related. We argued that this feature of the bulk absorption controlled
by a single parameter $m_4$ can be used to detect an extra dimension. In this paper, we showed that both of
these features depend on the presence of one or more extra dimensions, but not on the details of them,
in the sense that they do not depend on the details of the KK spectrum. With this finding, we
may now regard them to be generic features of strongly coupled extra dimensional models, and not just for the MM.

This research is supported by the Natural Sciences and Engineering Research
Council of Canada and by the Fonds de recherche sur la nature et les
technologies of Qu\,ebec.

\appendix
\section{}
Several formulas quoted in the text will be proved in this appendix.

To prove the formula \eq{qa} for $s_i=s(M_i)$, we start from its
definition in \eq{qt}:
\be
s_i=\sum_{a=1}^f{e_a^2\over(M_i-m_a)^2}\equiv\sum_{a=1}^fe_a^2x_{ai}^2.
\labels{siapp}\ee
Considered as a function of $M_i$, $s_i$ has a pole at every $m_a$. These 
are simple poles because $e_a^2$ has a zero at the same point.
The residue at $M_i=m_a$ can be obtained with the help of \eq{e2res} to be
$-\prod_{k\not=i}^{f-1}(-x_{ak})/\prod_{c\not=a}^fm_{ac}$. Moreover,
because of \eq{e2}, $s_i$ asymptotically appraches $1/M_i^2$ at large $M_i$.
The function on the right hand side of \eq{qa} has exactly the same
poles, the same residues, and the same asymptotic behavior, hence the 
difference between that function and $s_i$ is an entire function which vanishes
at infinity. This entire function must be zero, hence $s_i$ is given by
the function in \eq{qa}.

Next, we want to show the matrix $f\!\!\x\!\!f$ $V$ in \eq{v} to be an
orthogonal matrix.
We shall do that by showing each row of $V$ to be normalized, and 
two different rows to be mutually orthogonal.

Let us first prove that the $a$th row and the $b$th row are orthogonal, when
$a\not=b$. The dot product of these two rows is
\be
e_ae_b\[1+\sum_{i=1}^{f-1}{1\over (M_i-m_a)(M_i-m_b)s_i}\]
\equiv e_ae_b\[1+E\]\labels{orthoab}.\nn\ee
For this to be zero we must have $E=-1$. We will now show why this is so,
again using the complex variable theory.

Recall from \eq{qa} that
\be
s_i=-{\prod_{k\not=i}^{f-1}(M_i-M_k)\over\prod_{c=1}^f(M_i-m_c)}.\nn\ee
Hence
\be
E=-\sum_{i=1}^{f-1}{\prod_{c\not=a,b}^f(M_i-m_c)\over 
\prod_{k\not=i}^{f-1}(M_i-M_k)}\labels{ee}.\nn\ee
Consider $E$ as a function of $M_1$. It has a simple pole at 
$M_1=M_j$ for all $j>1$. Such a pole is contained in the $i=1$ term
and the $i=j$ term of the sum. The residue at $M_1=M_j$ coming from
the $i=1$ term is
$-\prod_{c\not=a,b}^f(M_j-m_c)/\prod_{k\not=1,j}^{f-1}(M_j-M_k)$,
and the residue coming from the $i=j$ term is exactly the opposite. Hence
the total residue is zero and $E$ is an entire function of $M_1$.
Asymptotically at large $M_1$, it follows from \eq{ee} that the only
contribution comes from the $i=1$ term, which gives $E=-1$,
hence $E$ is the constant function $-1$, which is required to show
the vanishing of \eq{orthoab} and the orthogonality of two different rows
of the matrix $V$.

Next, we show that every row of the matrix $V$ is normalized, namely,
\be
 e_a^2\[1+\sum_{i=1}^{f-1}{1\over (M_i-m_a)^2s_i}\]=1.\labels{normrow}\ee
We will show this in the form
\be
1+\sum_{i=1}^{f-1}{1\over (M_i-m_a)^2s_i}=
{1\over e_a^2}.\labels{normrow1}\ee
Using \eq{e2res}, \eq{qa}, this is equivalent to showing that
\be
1-\sum_{i=1}^{f-1}{1\over x_{ai}}{\prod_{c\not=a}^{f}x_{ci}\over \prod_{k\not=i}^{f-1}M_{ik}}
={\prod_{c\not=a}^fm_{ac}\over\prod_{k=1}^{f-1}(-x_{ak})}.\labels{normrow2}\ee

Let us label the function on the left as $L(m_a)$, and the function on
the right as $R(m_a)$. The function
 $L(m_a)$ goes to 1 for large $m_a$, and so does $R(m_a)$.
Moreover, $L(m_a)$ has
a simple pole at $m_a=\l_i$ for every $i$, with residue 
$\prod_{c\not=a}^fx_{ci}/\prod_{k\not=i}^{f-1}M_{ik}$. 
The function $R(m_a)$ also has a simple pole at $m_a=\l_i$ for every $i$,
with the same residue. Hence $L(m_a)-R(m_a)$ is an entire function that 
vanishes at infinity, so it is the constant function 0. This proves
\eq{normrow2}, and hence the normalization condition \eq{normrow}.

We have thus shown that the matrix $V$ is a real orthogonal matrix.

\end{document}